\documentclass[showkeys,12pt, preprint,preprintnumbers,nofootinbib, groupedaddress,superscriptaddress,amsmath,amssymb]{revtex4}
\usepackage{graphicx}
\usepackage{dcolumn}
\usepackage{bm}
\usepackage{amssymb}
\usepackage{amsmath}
\usepackage{epsfig}    
\usepackage{color}
\usepackage{slashed}
\usepackage{hhline}

\def\be{\begin{equation}}
\def\ee{\end{equation}}
\newcommand{\bea}{\begin{eqnarray}}
\newcommand{\eea}{\end{eqnarray}}

\numberwithin{equation}{section}

\begin{document}

    \rightline{KIAS-P20036, APCTP Pre2020-010}

\title{Dark matter physics in dark $SU(2)$ gauge symmetry \\ with non-Abelian kinetic mixing } 
\author{ P. Ko}
\email{pko@kias.re.kr}
\affiliation{School of Physics, KIAS, Seoul 02455, Korea}
\affiliation{Quantum Universe Center, KIAS, Seoul 02455, Korea}

\author{Takaaki Nomura}
\email{nomura@kias.re.kr}
\affiliation{School of Physics, KIAS, Seoul 02455, Korea}

\author{Hiroshi Okada}
\email{hiroshi.okada@apctp.org}
\affiliation{Asia Pacific Center for Theoretical Physics (APCTP) - Headquarters San 31, Hyoja-dong, Nam-gu, Pohang 790-784, Korea}

\date{\today}

\begin{abstract}

We investigate a model of dark sector based on non-Abelian $SU(2)_D$ gauge symmetry. 
This dark gauge symmetry is broken into discrete $Z_2$ via vacuum expectation values of 
two real triplet scalars,  and an $SU(2)_D$ doublet Dirac fermion becomes $Z_2-$odd particles 
whose lighter component makes stable dark matter candidate.
The standard model and dark sector can be connected via the scalar mixing and the gauge kinetic 
mixing generated by higher dimensional operators. 
We then discuss relic density of dark matter and implications to collider physics in the model.
{The most unique signatures of this model at the LHC would be the dark scalar ($\Phi_1^{(')}$) 
productions where it subsequently decays into : 
(1) a fermionic dark matter ($\chi_l$) and a heavy dark fermion ($\chi_h$) pair, $\Phi_1^{(')} \to \bar \chi_l \chi_h(\bar \chi_h \chi_l) $,
followed by $\chi_h$ decays into $\chi_l$ and non-Abelian dark gauge bosons ($X_i$'s) which decays into SM fermion pair $\bar f_{SM} f_{SM}$ 
resulting in the reaction $p p \rightarrow \Phi_1^{(')} \rightarrow \bar \chi_h \chi_l (\bar \chi_l \chi_h) \to f_{SM} \bar f_{SM} \chi_l \bar \chi_l $, 
(2) a pair of $X_i$'s followed by $X_i$ decays into a DM pair or the SM fermions resulting in the reaction, 
$p p \rightarrow \Phi_1^{(')} \rightarrow X_i X_i \rightarrow \bar \chi_l \chi_l f_{SM} \bar f_{SM}$ or even number of $f_{SM} \bar{f}_{SM}$ pairs. 
}
\end{abstract}
\maketitle
\newpage

\section{Introduction}

The standard model (SM) of particle physics has been very successful in describing phenomenology observed in various experiments.
However the existence of dark matter (DM) cannot be explained in the SM framework, 
and it would be described as a new particle associated with physics beyond the SM.
The nature of DM is an open question and there are many experimental searches for interactions among DM and the SM particles 
such as in direct detection, indirect detection and collider experiments.
No clear evidence of DM would indicate a dark sector which is hidden from current observations.

As the SM is described by local gauge symmetries, it is plausible that the dark sector is also ruled 
by a hidden/dark gauge symmetry.
Moreover stability of DM indicates necessity of a symmetry to protect it from decay, and it can be a remnant of dark gauge symmetry (see Ref.~\cite{Ko:2018qxz} for a review along this line). 
Thus, it is an attractive scenario that dark gauge symmetry is spontaneously broken 
to a symmetry stabilizing DM candidate. 
To realize this concept, we are especially interested in the extension of the SM introducing a new $SU(2)_D$ gauge symmetry where all the SM fields are singlet under it.  
The interesting properties of a model with local $SU(2)_D$ group is that an unbroken  discrete symmetry can be naturally preserved  after the spontaneous breaking of the $SU(2)_D$ gauge symmetry 
comparing with a hidden local $U(1)$ case in which the $U(1)$ charge has to satisfy some artificial tuning~\cite{Krauss:1988zc}.
Note that  various applications of the hidden $SU(2)$ gauge symmetry have been studied in literatures, for examples, a remaining $Z_3(Z_4)$ symmetry with a quadruplet(quintet) in ref.~\cite{Chiang:2013kqa, Chen:2015nea,Chen:2015dea}, $Z_2 \times Z'_2$ symmetry~\cite{Gross:2015cwa}, a custodial symmetry in refs.~\cite{Boehm:2014bia, Hambye:2008bq} and an unbroken $U(1)$ from $SU(2)$ in refs.~\cite{Baek:2013dwa,Khoze:2014woa,Daido:2019tbm}.

In this paper, we discuss a simple scenario in which $SU(2)_D$ gauge symmetry is broken into discrete $Z_2$ symmetry by vacuum expectation values (VEVs) of two $SU(2)_D$ triplet 
real scalar fields ($\vec \phi$ and $\vec\phi '$), and an $SU(2)_D$ doublet 
 Dirac fermion $\chi$ is $Z_2$-odd DM candidate.
We also introduce higher dimensional  operators that induce gauge kinetic mixing terms between 
$SU(2)_D$ and $U(1)_Y$ gauge fields after $SU(2)_D$ symmetry breaking by nonzero VEV's of 
$\vec \phi$ and $\vec \phi '$ as  mediators between the dark gauge sector and the SM sector.
After fixing our model, we formulate particle mass spectra and their interactions in the dark sector 
and the portals to the SM sector.
Then relic density of our DM candidate is estimated taking into account constraint from direct 
detection of DM.
Furthermore we discuss implications to collider physics considering the scalar portal and the kinetic 
mixing as connections between dark sector and the SM.

This paper is organized as follows.
In Sec.~II, we show our model of $SU(2)_D$ dark sector formulating mass spectra and interactions.
In Sec.~III, we analyze DM relic density and discuss the allowed parameter region.  
In Sec.~IV, we discuss implications to collider physics.
Finally We conclude and discuss in Sec.~V.

\section{Model and formulas}

 \begin{widetext}
\begin{center} 
\begin{table}[t]
\begin{tabular}{|c||c|c|c|c|}\hline\hline  
Fields & $\chi$ & $\vec \phi$ & $\vec \phi'$
\\\hline 
 $SU(2)_D$ & $\bf{2}$ & $\bf{3}$  & $\bf{3}$   \\\hline 
\end{tabular}
\caption{Charge assignment for the fields in $SU(2)_D$ dark sector where $\chi$ is Dirac fermion and $\vec \phi_i (i=1,2)$ are scalars.}
\label{tab:1}
\end{table}
\end{center}
\end{widetext}

In this section we summarize the setup for our model. 
We introduce a dark sector which is controlled by a non-Abelian $SU(2)_D$  dark gauge 
symmetry, with two real scalar fields $\vec \phi$ and $\vec \phi'$, and one Dirac fermion $\chi$ as summarized in Table~\ref{tab:1}. In components, $\vec \phi(\vec \phi')$ and $\chi$ are written as 
\begin{align}
\vec \phi \left[\vec \phi' \right] = (\phi_1[\phi'_1], \phi_2[\phi'_2], \phi_3[\phi'_3]), \quad \chi = ( \chi_1, \chi_2)^T, 
\end{align}
where the indices for triplet scalars correspond to three $SU(2)_D$ generators. 

The  $SU(2)_D$ dark gauge symmetry is spontaneously broken by nonzero VEV's of two real scalar 
triplets $\phi$ and $\phi'$.   In our scenario, we  assume 
VEV alignments of two scalar triplets as 
\begin{equation}
\langle \vec \phi \rangle = \left(0,0, \frac{v_\phi}{\sqrt{2}} \right), \quad \langle \vec \phi' \rangle = \left(\frac{v_{\phi'}}{\sqrt{2}}, 0, 0 \right).
\label{eq:alignment}
\end{equation}
When $SU(2)_D$ is broken by the VEVs of the triplets, the vacuum is invariant under 
the transformation defined by $U_{T_3} \equiv e^{i 2 \pi T_3 }$ as $U_{T_3} \langle \vec \phi^{(\prime)} \rangle = \langle  \vec \phi^{(\prime)} \rangle$, where $T_3$ is the diagonal component of $SU(2)_D$ generators. 
Then $\chi$ field transform $U_{T_3} \chi = - \chi$ since $T_3$ values of $\chi$'s 
components are $\pm 1/2$.
In general, we would obtain even (odd) parity under $U$ transformation for any component of 
$SU(2)_D$ multiplet with integer (half-integer) value of $T_3$.
Thus $SU(2)_D$ gauge symmetry is broken to $Z_2$ symmetry when we assume  
Eq. (\ref{eq:alignment}). 
Note that the $Z_2$ symmetry will guarantees the stability of DM candidate which is the lightest component with odd $Z_2$ parity.

Now we write down the Lagrangian for kinetic terms of the dark sector and the scalar potential: 
\begin{align}
\label{Eq:Lagrangian}
\mathcal{L}_D = & - \frac{1}{4} X^a_{\mu \nu} X^{a \mu \nu} +  D_\mu \vec \phi \cdot D^\mu \vec \phi +  D_\mu \vec \phi' \cdot D^\mu \vec \phi' + \bar \chi (D_\mu \gamma^\mu - M_\chi) \chi   \\ 
V  = &   \mu_H^2 H^\dagger H + \lambda_H (H^\dagger H)^2 \nonumber \\
& + \mu_1^2 \vec \phi \cdot \vec \phi + \mu_2^2 \vec \phi' \cdot \vec \phi'
 + \lambda_1 \left( \vec \phi \cdot \vec \phi \right)^2 + \lambda_2  \left( \vec \phi' \cdot \vec \phi' \right)^2
+ \lambda_{3} \left( \vec \phi \cdot \vec \phi' \right)^2 \nonumber \\
& + \lambda_{4}  \left( \vec \phi \cdot \vec \phi \right)  \left(\vec \phi \cdot \vec \phi' \right)
+ \lambda_{5} \left( \vec \phi' \cdot \vec \phi' \right)  \left(\vec \phi \cdot \vec \phi' \right) + \lambda_{6}  \left( \vec \phi \cdot \vec \phi \right)   \left( \vec \phi' \cdot \vec \phi' \right)  \nonumber \\
&+ \lambda_{H\phi}  \left( \vec \phi \cdot \vec \phi \right)  \left( H^\dagger H \right)
+\lambda_{H\phi'}  \left( \vec \phi' \cdot \vec \phi' \right)  \left( H^\dagger H  \right) \nonumber \\
& + \frac{y_{\chi \phi}}{2} \bar \chi \left( \vec \phi \cdot \vec \sigma \right) \chi + \frac{y_{\chi \phi'}}{2} \bar \chi \left( \vec \phi' \cdot \vec \sigma \right) \chi 
\label{eq:potential}
\end{align}
where $X_{\mu \nu}^a (a=1,2,3)$ is the field strength of $SU(2)_D$ gauge field, and $H$ is the 
SM Higgs doublet field  written as 
\begin{equation}
H = \begin{pmatrix} G^+ \\ \frac{1}{\sqrt{2}} (v + \tilde h + i G_Z) \end{pmatrix} . 
\end{equation}
Here $v_H$ is the VEV of the SM Higgs doublet, $H$, and $G^\pm$ and $G_{Z}$ are Nambu-Goldstone(NG) bosons absorbed by $W^\pm$ and $Z$ bosons.

\subsection{Scalar sector}

Here we consider the scalar sector of the model.
Firstly we consider conditions to get VEV alignment in Eq.~\eqref{eq:alignment}.
From the stationary conditions $\partial V/\partial \phi_i = 0$ and $\partial V/\partial \phi'_i = 0$, we obtain following non-trivial conditions (or vanishing tadpole conditions): 
\begin{align}
& \lambda_1 v_{\phi}^3 + \frac{1}{2} \lambda_6 v_{\phi} v^2_{\phi'} + \frac{1}{2} \lambda_{H \phi } v_{\phi} v_H^2 - \mu_1^2 v_\phi = 0, \nonumber  \\
& \lambda_2 v_{\phi'}^3 + \frac{1}{2} \lambda_6 v_{\phi}^2 v_{\phi'} + \frac{1}{2} v_{\phi} \lambda_{H \phi'} v_H^2 - \mu_2^2 v_{\phi'} = 0, \nonumber \\
& \lambda_H v_H^3 + \frac{1}{2} \lambda_{H \phi} v^2_\phi v_H + \frac{1}{2} \lambda_{H\phi'} v^2_{\phi'} v_H - \mu_H^2 v_H = 0, \nonumber \\
& \lambda_4 v_{\phi}^2 + \lambda_5 v_{\phi'}^2 =0. \label{eq:VEVconditions}
\end{align}
The mass terms of scalar fields are given  by the quadratic terms in the scalar fields in the Lagrangian: 
\begin{align}
{\color{blue} - }
\mathcal{L}_{M_S} = & \frac{1}{4} \lambda_3 v^2_{\phi'} \phi_1^2 + \frac{1}{4} \lambda_4 v_\phi v_{\phi'} \phi_1 \phi_3 - \frac{1}{2} \lambda_4 v_\phi^2 \phi_1 \phi'_1
+ \frac{1}{2} \lambda_3 v_{\phi} v_{\phi'} \phi_1 \phi'_3 + \lambda_1 v^2_{\phi} \phi_3^2 \nonumber \\
& + \lambda_6 v_\phi v_{\phi'} \phi_3 \phi'_1 + \frac{1}{2} \lambda_4 v^2_{\phi} \phi_3 \phi'_3 +  \lambda_2 v^2_{\phi'} \phi'^2_1 - \frac{1}{2}  \lambda_4 \frac{ v^3_{\phi} }{v_\phi'} \phi'_1 \phi'_3 
+ \frac{1}{4} \lambda_3 v^2_{\phi} \phi'^2_3 \nonumber \\
& + \lambda_H v_H^2 \tilde h^2 +  \lambda_{H \phi} v_{\phi} v_H \phi_3 \tilde h +  \lambda_{H \phi'} v_{\phi'} v_H \phi'_1 \tilde h,
\label{eq:Smass1}
\end{align}
where we used the last equation of Eq.~\eqref{eq:VEVconditions} to substitute $\lambda_5$.
Notice that mass terms associated with $\phi_2$ and $\phi'_2$ are absent and they are identified as Nambu--Goldstone(NG) bosons which are absorbed by the two massive gauge bosons in 
the dark sector.  

From now on, we shall assume $\lambda_4 \ll 1$ to simplify the scalar mass terms. 
Then Eq.~(\ref{eq:Smass1}) becomes
\begin{align}
{\color{blue} - }
\mathcal{L}_{M_S} \simeq & \frac{1}{4} \lambda_3 v^2_{\phi'} \phi_1^2 + \frac{1}{2} \lambda_3 v_{\phi} v_{\phi'} \phi_1 \phi'_3 + \lambda_1 v^2_{\phi} \phi_3^2 
 + \lambda_6 v_\phi v_{\phi'} \phi_3 \phi'_1  +  \lambda_2 v^2_{\phi'} \phi'^2_1 + \frac{1}{4} \lambda_3 v^2_{\phi} \phi'^2_3 \nonumber \\
& + \lambda_H v_H^2 \tilde h^2 +  \lambda_{H \phi} v_{\phi} v_H \phi_3 h +  \lambda_{H \phi'} v_{\phi'} v_H \phi'_1 h,
\end{align}
ignoring terms  with the $\lambda_4$ coupling.
The terms for $\phi_1$ and $\phi'_3$ can be organized as 
\begin{equation}
\frac{1}{4} \lambda_3 (v_\phi^2 + v_{\phi'}^2) \left( \cos \alpha \phi_1 + \sin \alpha \phi'_3 \right)^2,
\end{equation}
where $\sin \alpha (\cos \alpha) = v_{\phi'}(v_\phi)/\sqrt{v_\phi^2 + v_{\phi'}^2}$.
We then find that the mass eigenstate $( \cos \alpha \phi_1 + \sin \alpha \phi'_3 )$ has the 
mass eigenvalue $\lambda_3 (v_\phi^2 + v_{\phi'}^2)/2$, whereas its orthogonal state 
$(-\sin \alpha \phi_1 + \cos \alpha \phi'_3)$ corresponds to the NG boson absorbed by $SU(2)_D$ 
dark gauge boson. 

Finally,  the   mass matrix for $(h, \phi_3, \phi'_1)$ is given by
\begin{equation}
{- } \mathcal{L}_{M_S} \supset 
\frac12 \begin{pmatrix} h \\ \phi_3 \\ \phi'_1  \end{pmatrix}^T
\begin{pmatrix} 2 \lambda_H v_H^2 & \lambda_{H \phi} v_\phi v_H & \lambda_{H \phi'} v_{\phi'} v_H \\  \lambda_{H \phi} v_\phi v_H & 2 \lambda_1 v_\phi^2 & \lambda_6 v_\phi v_{\phi'} \\ 
\lambda_{H \phi'} v_{\phi'} v_H &  \lambda_6 v_\phi v_{\phi'} & 2 \lambda_2 v_{\phi'}^2 \end{pmatrix} 
\begin{pmatrix} h \\ \phi_3 \\ \phi'_1  \end{pmatrix}.
\end{equation}
Thus $\phi_3$ and $\phi'_1$ can mix with the SM Higgs field and interact with SM particle via mixing effects.
In our phenomenological analysis, we discuss the following two simplified cases. \\
{\bf Scenario (1)}: $\lambda_{H \phi'}, \lambda_6 \to 0$ \\
In this case, $\tilde h$ and $\phi_3$ mix while $\phi'_1$ is almost the mass eigenstate without mixing.
Then squared mass terms for $\{ \tilde h, \phi_3 \}$ are given by
\begin{equation}
{ - } 
\mathcal{L} \supset \frac{1}{2} \begin{pmatrix} \tilde h \\ \phi_3 \end{pmatrix}^T \begin{pmatrix} 2 \lambda_H v_H^2 & \lambda_{H \phi} v_\phi v_H \\  \lambda_{H \phi} v_\phi v_H  & 2 \lambda_1 v^2_\phi \end{pmatrix} \begin{pmatrix} \tilde h \\ \phi_3 \end{pmatrix}.
\end{equation} 
This squared mass matrix can be diagonalized by an orthogonal matrix,  and the resulting mass eigenvalues are given by 
\begin{equation}
m_{h,\Phi_1}^2 =   \lambda_H v_H^2 + \lambda_1 v_\phi^2  \pm \sqrt{\left(  \lambda_H v_H^2 -  \lambda_1 v_\phi^2 \right)^2 +  \lambda_{H \phi}^2 v_\phi^2 v_H^2 } \ .
\end{equation}
The relevant mass eigenstates $h$ and $\Phi_1$ are also given by   
\begin{equation}
\begin{pmatrix} h \\ \Phi_1 \end{pmatrix} = \begin{pmatrix} \cos \alpha & \sin \alpha \\ - \sin \alpha & \cos \alpha \end{pmatrix} \begin{pmatrix} \tilde h \\ \phi_3 \end{pmatrix}, \quad
\tan 2 \alpha = \frac{ \lambda_{H \varphi} v_\phi v_H}{ \lambda_H v_H^2 -  \lambda_1 v_\phi^2}, \label{eq:massE1}
\end{equation}
where $\alpha$ is the mixing angle, and $h$ is identified as the SM-like Higgs boson.
Also we rewrite $\phi'_1$ as an approximate mass eigenstate such that
\begin{equation}
\Phi_2 \simeq \phi'_1, \quad m_{\Phi_2}^2 \simeq 2 \lambda_2 v^2_{\phi'} \ . \label{eq:massE2}
\end{equation}
The scalar mixing is constrained by Higgs precision measurements as 
$\sin \alpha \lesssim 0.3$ when the SM Higgs does not decay into particles in the dark sector~\cite{Chpoi:2013wga, Cheung:2015dta}
We investigate the bound for $\sin \alpha$ when the SM Higgs decays into dark gauge bosons below.  \\
%
{\bf Scenario (2)}: $\lambda_{H \phi}, \lambda_6 \to 0$ \\
In this case, we obtain mass eigenvalues and eigenstates by replacement $\lambda_{H\phi} \to \lambda_{H \phi'}$, $v_{\phi} \to v_{\phi'}$, $\lambda_1 \leftrightarrow \lambda_2$ 
and $\phi_3 \leftrightarrow \phi'_1$ for case (1).
Then they are given by
\begin{align}
& m_{h,\Phi'_1}^2 =  \lambda_H v_H^2 + \lambda_2 v_{\phi'}^2  \pm  \sqrt{\left(  \lambda_H v_H^2 -  \lambda_2 v_{\phi'}^2 \right)^2 +  \lambda_{H \phi'}^2 v_{\phi'}^2 v_H^2 }, \\
& \begin{pmatrix} h \\ \Phi'_1 \end{pmatrix} = \begin{pmatrix} \cos \alpha' & \sin \alpha' \\ - \sin \alpha' & \cos \alpha' \end{pmatrix} \begin{pmatrix} \tilde h \\ \phi'_1 \end{pmatrix}, \quad
 \tan 2 \alpha' = \frac{ \lambda_{H \varphi'} v_{\phi'} v_H}{ \lambda_H v_H^2 -  \lambda_2 v_{\phi'}^2}, \label{eq:massE3}
\end{align}
where $\alpha'$ is the mixing angle, and $h$ is again identified as the SM-like Higgs boson.
Also we rewrite $\phi_3$ as approximate mass eigenstate such that
\begin{equation}
\Phi'_2 \simeq \phi_3, \quad m_{\Phi'_2}^2 \simeq 2 \lambda_1 v^2_{\phi}. \label{eq:massE4}
\end{equation}

\subsection{Gauge sector}

The dark and the SM sectors can interact through terms in the potential associated with the SM Higgs in Eq.~(\ref{eq:potential}) that is called the  Higgs portal.
In addition the dark gauge sector and the SM gauge sector can be concocted via kinetic mixings  
between $SU(2)_D$ and $U(1)_Y$ after $SU(2)_D$ gauge symmetry breaking by nonzero 
VEVs of $\vec \phi$ and $\vec \phi '$~\cite{Arguelles:2016ney}.
The relevant terms for these kinetic mixings  are two dim-5 operators:
\begin{align}
\mathcal{L}_{XB} = & \frac{C_\phi}{\Lambda} X^a_{\mu \nu} B^{\mu \nu} \phi^a + \frac{C_{\phi'}}{\Lambda} X^a_{\mu \nu} B^{\mu \nu} \phi'^a,
\end{align}
where $\Lambda$ indicate the cut off scale and $B^{\mu \nu}$ is the gauge field strength for $U(1)_Y$.
After $\phi$ and $\phi'$ developing VEVs, we obtain the following kinetic mixing terms: 
\begin{equation}
\mathcal{L}_{\rm KM} = - \frac{1}{2} \sin {\delta_1} X^1_{\mu \nu} B^{\mu \nu} - \frac{1}{2} \sin {\delta_3} X^3_{\mu \nu} B^{\mu \nu},
\label{eq:kinetic_mixing}
\end{equation}
where we defined $\sin {\delta_1} \equiv \sqrt{2} C_{\phi'} v_\phi/\Lambda$ and $\sin { \delta_3} \equiv \sqrt{2} C_{\phi} v_{\phi'}/\Lambda$ as new kinetic mixing parameters.

The kinetic terms for $X^{1,3}_\mu$ and $B_\mu$ can be diagonalized by the following transformations: 
\begin{align}
& B_\mu = \tilde{B}_\mu - \tan { \delta_1} \tilde{X}^1_\mu - \frac{1}{\cos {\delta}} (\tan {\delta_3} - \tan { \delta_1} \sin { \delta}) \tilde{X}^3_\mu, \\
& X^1_\mu = \frac{1}{\cos { \delta_1}} \tilde{X}^1_\mu - \frac{\tan { \delta}}{\cos {\delta_1}} X^3_\mu, \\
& X^3_\mu = \frac{1}{\cos { \delta_3} \cos { \delta}} \tilde{X}^3_\mu,
\end{align}
where ${\delta}$ is defined as $\sin { \delta} \equiv - \tan {\delta_1} \tan { \delta_3}$.
In our analysis, we take a limit of ${\delta_1} \ll 1$ and ${\delta_3} \ll 1$ and gauge fields are approximately written by
\begin{equation}
B \simeq \tilde{B} - { \delta_1} X^1_\mu - {\delta_3} X^3_\mu, \quad X^1_\mu \simeq \tilde X^1_\mu, \quad X^3_\mu \simeq \tilde{X}^3_\mu.
\end{equation}
Then we denote dark gauge bosons associated with $X^{1,2,3}_\mu$ field as $X_{1,2,3}$ henceforth.
Note that mixing with $Z$ boson is suppressed unless dark gauge boson and the SM $Z$ boson 
masses are not close enough. 
In our analysis, we assume a dark gauge boson  mainly mixes with photon field. 

After two triplet scalar fields develop nonzero VEVs, $SU(2)_D$ gauge bosons obtain masses 
from kinetic term such that
\begin{align}
\mathcal{L}_M =  g_D^2 v_\phi^2 X^1_\mu X^{1 \mu} +  g^2_D (v_{\phi}^2 + v^2_{\phi'}) X^2_\mu X^{2 \mu} +  g_D^2 v_{\phi'}^2 X^3_\mu X^{3 \mu} .
\end{align}
Here we have ignored kinetic mixing effects since it is negligibly small in our scenario.
We thus find the masses of dark gauge bosons  be 
\begin{equation}
m_{X_1} = \sqrt{2} g_D v_\phi, \quad  m_{X_2} 
= g_D \sqrt{2(v_\phi^2 + v^2_{\phi'})}, \quad m_{X_3} = \sqrt{2} g_D v_{\phi'} .
\end{equation}
Note that the $X_2$ is always the heaviest one.
In addition, three-point interactions among scalar and gauge bosons are given by
\begin{equation}
\mathcal{L} \supset g_D^2 v_\phi \phi_3 (X^1_\mu X^{1 \mu} + X^2_\mu X^{2 \mu}) + g_D^2 v_{\phi'} \phi'_1 (X^2_\mu X^{2 \mu}+ X^3_\mu X^{3 \mu}),
\end{equation}
where $\phi_3$ and $\phi'_1$ can be written as mass eigenstates using Eqs.~(\ref{eq:massE1}) and (\ref{eq:massE2}) for  the case (1) and using Eqs~(\ref{eq:massE3}) and (\ref{eq:massE4}) for the 
case (2) described in previous subsection.

Finally interactions among dark gauge fields are also written $\rightarrow$ given by
\begin{equation}
\mathcal{L} \supset - g_D \epsilon^{abc} \partial_\mu X^a_\nu X^{b \mu} X^{c \nu} - \frac14 g_D^2 \epsilon^{abc} \epsilon^{ade} X^b_\mu X^c_\nu X^{d \mu} X^{e \nu},
\end{equation}
where $\epsilon^{abc}$ is the structure constants of $SU(2)_D$  and $a = 1,2,3$.
The heaviest gauge boson $X_2$ would decay into $X_1 X_3$ through the three point 
gauge interaction, where the $X_1$ and/or {$X_3$} transition 
will be off-shell 
due to the mass relation among dark gauge bosons and both of them will eventually decay into 
the SM particles through kinetic mixings, Eq. (\ref{eq:kinetic_mixing}). 

\subsection{Fermions in the  dark sector}

The mass terms of $SU(2)_D$ doublet fermion are given by
\begin{align}
{\color{blue} -} \mathcal{L} & 
\supset M_\chi (\bar \chi_1 \chi_1 + \bar \chi_2 \chi_2) + \frac{y_{\chi \phi} v_\phi}{2} (\bar \chi_1 \chi_1 - \bar \chi_2 \chi_2) + \frac{y_{\chi \phi'} v_{\phi'}}{2} (\bar \chi_1 \chi_2 + \bar \chi_2 \chi_1) \nonumber \\
& \equiv M_{11} \bar \chi_1 \bar \chi_1 + M_{12}(\bar \chi_1 \chi_2 + \bar \chi_2 \chi_1) + M_{22} \bar \chi_2 \chi_2,
\end{align}
where we assumed all coefficients are real.  The mass splitting and the mass mixings 
between $\chi_1$ and $\chi_2$ are induced by the $y_{\chi \phi}$ and $y_{\chi \phi '}$ respectively, 
in the last line of Eq. (\ref{eq:potential}).
The mass eigenvalues and eigenstates are  obtained in a straight forward manner  as
\begin{align}
& m_{\chi_l, \chi_h} = \frac{1}{2} (M_{11}+M_{22}) \pm \frac{1}{2} \sqrt{(M_{11}- M_{22})^2 + 4 M_{12}^2}, \nonumber \\
& \left( \begin{array}{c} \chi_1 \\ \chi_2 \end{array} \right) = 
\left( \begin{array}{cc} \cos \theta_\chi & - \sin \theta_\chi \\ \sin \theta_\chi & \cos \theta_\chi \end{array} \right)
\left( \begin{array}{c} \chi_l \\ \chi_h \end{array} \right),
\end{align}
where $m_{\chi_l} < m_{\chi_h}$  by definition.  
The mixing angle $\theta_\chi$  is given by
\begin{equation}
\tan 2 \theta_\chi = \frac{2 M_{12}}{M_{11} - M_{22}} = \frac{y_{\chi \phi'} v_{\phi'}}{y_{\chi \phi} v_\phi}.
\end{equation}
Furthermore we obtain interactions among scalar fields and mass eigenstates of dark fermions such that
\begin{align}
\mathcal{L} \supset & \frac{y_{\chi \phi}}{2} \Bigl[ \phi_3 \left( \cos 2 \theta_\chi \bar \chi_l \chi_l - \cos 2 \theta_\chi \bar \chi_h \chi_h - \sin 2 \theta_\chi (\bar \chi_l \chi_h + \bar \chi_h \chi_l) \right)   
\nonumber \\
& \qquad + \phi_1 \left( \sin 2 \theta_\chi \bar \chi_l \chi_l - \sin 2 \theta_\chi \bar \chi_h \chi_h + \cos 2 \theta_\chi (\bar \chi_l \chi_h + \bar \chi_h \chi_l) \right) \Bigr] \nonumber \\
& + \frac{y_{\chi \phi'}}{2} \Bigl[ \phi'_3 \left( \cos 2 \theta_\chi \bar \chi_l \chi_h - \cos 2 \theta_\chi \bar \chi_h \chi_h - \sin 2 \theta_\chi (\bar \chi_l \chi_h + \bar \chi_h \chi_l) \right)  \nonumber \\
& \qquad \qquad + \phi'_1 \left( \sin 2 \theta_\chi \bar \chi_l \chi_l - \sin 2 \theta_\chi \bar \chi_h 
\chi_h + \cos 2 \theta_\chi (\bar \chi_l \chi_h + \bar \chi_h \chi_l) \right) \Bigr],
\end{align}
where $\phi_{1,3}$ and $\phi'_{1,3}$ are substituted to mass eigenstates as discussed in previous subsection.

{
\subsection{Topological $Z_2$ string}

In our DM model, the original non-Abelian gauge symmetry $G \equiv SU(2)_D$ is spontaneously 
broken down to its subgroup $H = Z_2 =\{1,-1\}$ which is disconnected.  
Then the vacuum manifold of the model is given by ${\cal M} = G/H $ so that the first homotopy 
group of ${\cal M}$ is $\pi_1 (G/H) = \pi_0 (H) = H = Z_2$ \cite{Weinberg:2012pjx}. 
Therefore in the particle spectra of this model, there will be $Z_2$ string which is a topological object. 
One $Z_2$ vortex is topologically nontrivial, but two of them can be deformed smoothly into 
the vacuum, thereby being topologically trivial.   These $Z_2$ string can contribute to the dark matter
of the current Universe to some extent, but detailed study of this issue is beyond the scope of 
this paper.  In the following, we shall simply ignore topological $Z_2$ strings assuming their 
contribution to the Universe is negligible.
}

\section{Dark matter}

In this section, we discuss DM phenomenology in our model including DM relic density.
In our scenario, DM is the lightest  lightest dark fermion $\chi_l$ which is stabilized 
by the remnant dark $Z_2$ symmetry after dark gauge symmetry breaking.
Firstly we require that Higgs portal interactions of DM are suppressed in order to avoid severe 
constraints from DM direct detection experiments.
For the scenario (1) of Higgs mixing, we prefer large mixing of dark fermions, 
$\theta_\chi \sim \pi/4$, since DM couples with Higgs via $\phi_3$.
On the other hand, for the scenario (2), we prefer small mixing, $|\theta_\chi| << 1$, 
since DM couples with Higgs via $\phi'_1$.  
Then relic density of DM is determined by gauge interactions in dark sector in our scenarios 
where we assume dark scalars are heavier than DM.

Then, the relevant interaction terms are 
\begin{align}
\mathcal{L}  \supset & \frac{g_D}{2} \left( \cos 2 \theta_\chi \bar \chi_l \gamma^\mu \chi_l - \sin 2 \theta_\chi \bar \chi_l \gamma^\mu \chi_h
 - \sin 2 \theta_\chi \bar \chi_h \gamma^\mu \chi_l  - \cos 2 \theta_\chi \bar \chi_h \gamma^\mu \chi_h \right) X^3_\mu \nonumber \\
 & + \frac{g_D}{2} \left( \sin 2 \theta_\chi \bar \chi_l \gamma^\mu \chi_l + \cos 2 \theta_\chi \bar \chi_l \gamma^\mu \chi_h
 + \cos 2 \theta_\chi \bar \chi_h \gamma^\mu \chi_l  - \sin 2 \theta_\chi \bar \chi_h \gamma^\mu \chi_h \right) X^1_\mu \nonumber \\
 & + i \frac{g_D}{2} \left( \bar \chi_h \gamma^\mu \chi_l - \bar \chi_l \gamma^\mu \chi_h \right) X^2_\mu + e c_W \delta_1 X^1_\mu J^\mu_{EM} + e c_W \delta_3 X^3_\mu J^\mu_{EM}, \\
 J_{EM}^{\mu} & = \sum_{f_{SM}} Q_{f_{SM}} \bar f_{SM} \gamma^\mu f_{SM},
\end{align}
where $J_{EM}^{\mu}$ is electromagnetic current and $Q_{f_{SM}}$ is the electric charge of the SM fermions $f_{SM}$.
Then we implement these interactions in {\tt micrOMEGAs 4.3.5}~\cite{Belanger:2014vza} to estimate relic density.

In our analysis we take the dark fermion mixing angle for each scenario as 
\begin{align}
& \theta_\chi = \frac{\pi}{4} \quad \text{for the scenario (1)}, \\
& \theta_\chi = 0 \quad \text{for the scenario (2)},  
\end{align}
in order to avoid direct detection constraint.
For illustration, we consider two benchmark cases of dark gauge boson masses for each scenario;
\begin{align}
& \text{Scenario (1)}: m_{X_{1}} < m_{X_{3}}, \\
& \text{Scenario (2)}:  m_{X_{1}} > m_{X_{3}}, 
\end{align}
and $m_{X_2} = \sqrt{m^2_{X_1} + m^2_{X_3}}$.  
Also we require interaction between DM and the lightest dark gauge boson not to be suppressed 
by the dark fermion mixing effect.
In the following, we shall  focus on the scenario (1) since we just obtain similar results by replacing 
the role of $X_1$ and $X_3$ for the scenario (2).

In addition, we take into account $DM$-nucleon scattering via $Z'$ boson exchanging process.
The cross section for this process is calculated in non-relativistic limit as 
\begin{equation}
\sigma \simeq \frac{\delta_1^2 g_D^4}{2 \pi} \frac{1}{m_{X_1}^4} \left( \frac{m_{\chi_l} m_p}{m_{\chi_l} + m_{p}} \right)^2,
\end{equation}
where $m_p$ indicates the proton mass. 
Since  the $Z'$-SM fermion interaction is proportional to electromagnetic current, the DM  scattering 
with neutron is suppressed. 
For $m_{\chi_l} = \mathcal{O}(100)$ GeV, we obtain 
\begin{equation}
\sigma \sim 6 \delta_1^2 g_D^4 \left( \frac{100 \ {\rm GeV}}{m_{X_1}} \right)^4 \times 10^{-37} \ {\rm cm}^2.
\end{equation}
We then assume $\delta_1 \lesssim 10^{-5}$ to avoid direct detection constraints such as XENON1T~\cite{Aprile:2018dbl} and PandaX-II~\cite{Cui:2017nnn} 
which provide upper limit of $\sim 10^{-46}$ cm$^2$ for DM mass of $\sim 100$ GeV.

\begin{figure}[tb]\begin{center}
\includegraphics[width=80mm]{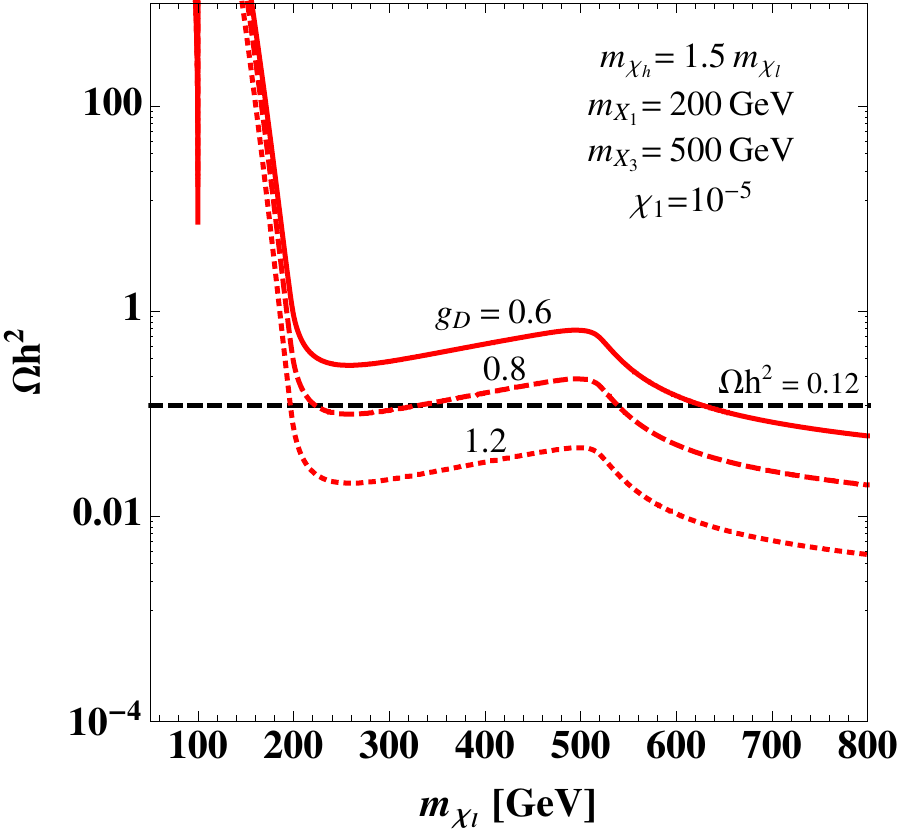} \
\includegraphics[width=80mm]{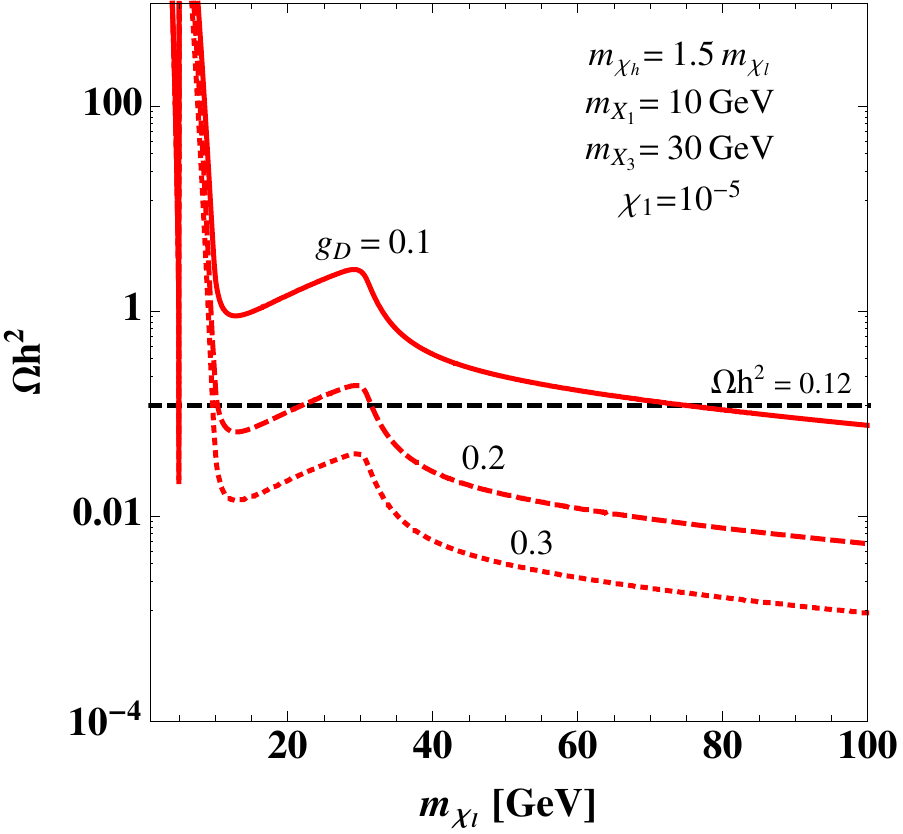} \
\caption{Relic density as a function of DM mass where relevant parameters are fixed as indicated on the plots.}   
\label{fig:DM1}\end{center}\end{figure}
In Fig.~\ref{fig:DM1}, we show thermal relic density of DM,  adopting dark gauge boson masses 
$\{m_{X_1}, m_{X_3} \} $ as $\{200, 500 \}$ GeV and $\{10, 30 \}$ GeV as reference values, 
$m_{\chi_h} = 1.5 m_{\chi_l}$, $\delta_{1,3} = 10^{-5}$, and some relevant values of 
gauge coupling $g_D$. 
We find that relic density is decreased when $\chi_l \bar \chi_l \to X_1 X_1 $ and 
$\chi_l \bar \chi_l \to X_2 X_2$ processes are kinematically allowed.
Then larger gauge coupling is required for larger $m_{X_1}$ mass to accommodate with observed 
thermal relic density of DM.
We can also explain relic density around resonance $2 m_{X_{\chi_l}} \sim m_{X_1}$ when 
$X_1$ mass is relatively light while the relic density tends to be larger than observed one for 
heavier dark gauge boson case due to small kinetic mixing parameter.

\begin{figure}[tb]\begin{center}
\includegraphics[width=80mm]{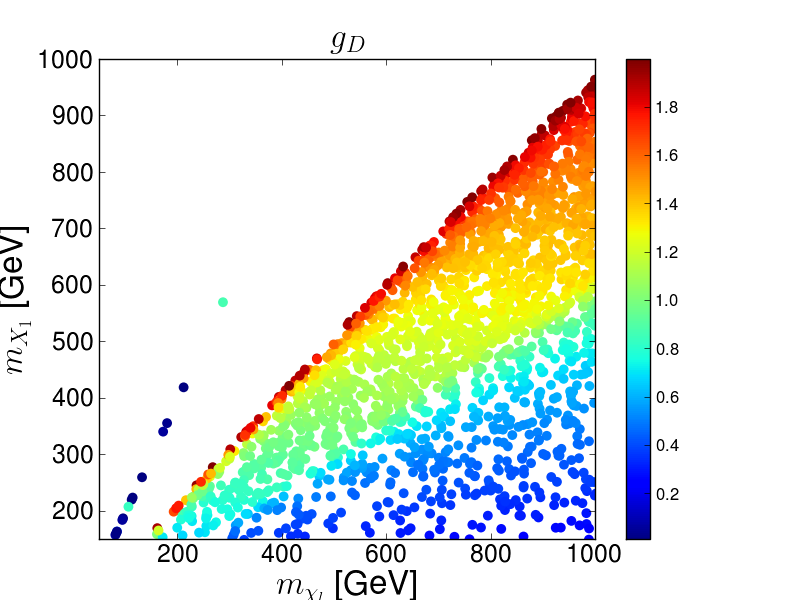} \
\includegraphics[width=80mm]{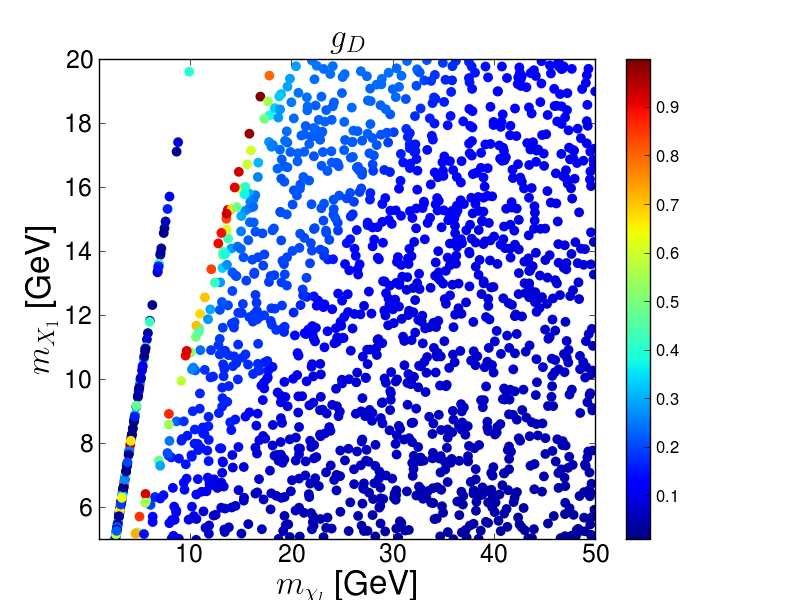} \
\caption{Parameter region satisfying relic density of DM.}   
\label{fig:DM2}\end{center}\end{figure}

Next we scan free parameters fixing ${\delta_{1,3}} = 10^{-5}$ to avoid direct detection constraint.
The two region for masses of $\chi_{l,h}$ and $X_{1,3}$ are considered in the range of
\begin{align}
 \text{Region I \ :} & \quad m_{\chi_l} \in [1, 50] {\rm GeV}, \quad m_{\chi_h} = 1.5 m_{\chi_l}, 
 \nonumber \\
& \quad m_{X_1} \in [5, 20] {\rm GeV}, \quad m_{X_3} = 1.5 m_{X_1}, \quad g_D \supset [0.01,1], \\
\text{Region II :} & \quad m_{\chi_l} \in [50, 1000] {\rm GeV}, \quad m_{\chi_h} = 1.5 m_{\chi_l}, 
\nonumber \\
& \quad m_{X_1} \in [150, 1000] {\rm GeV}, \quad m_{X_3} = 1.5 m_{X_1}, \quad g_D \supset [0.05, 2], 
\end{align}
where masses of $\chi_h$ and $X_{3}$ are respectively determined by those of $\chi_l$ and $X_1$ for simplicity.
We then search for the parameter region which provide observed DM thermal relic density 
approximately in the range of $0.11 < \Omega h^2 < 0.13$.  
In the left and right panels of Fig.~\ref{fig:DM2}, we show allowed parameter region on 
$\{m_{DM}, m_{X_1} \}$ plane for the region I and II where color gradient indicates values of $g_D$. 
We find large allowed region when $\chi_l \bar \chi_l \to X_1 X_1 (X_2 X_2)$ processes are kinematically allowed.
On the other hand, for $m_{X_1} > m_{\chi_l}$, we need some fine tuning around 
$m_{X_1} \simeq 2m_{\chi_l}$ to obtain resonant enhancement of annihilation cross section. 

\section{Implications in collider physics}

Here we discuss collider physics in the model focusing extra scalar boson production at the LHC via Higgs mixing.
Then extra scalar boson decays into dark gauge boson, dark fermion or SM particles where the branching ratio (BR) depend on parameters in the model.
We then discuss possible signals of the model at the LHC.

\subsection{Constraint from the SM Higgs boson decay \label{sec:HiggsConst}}

\begin{figure}[tb]\begin{center}
\includegraphics[width=80mm]{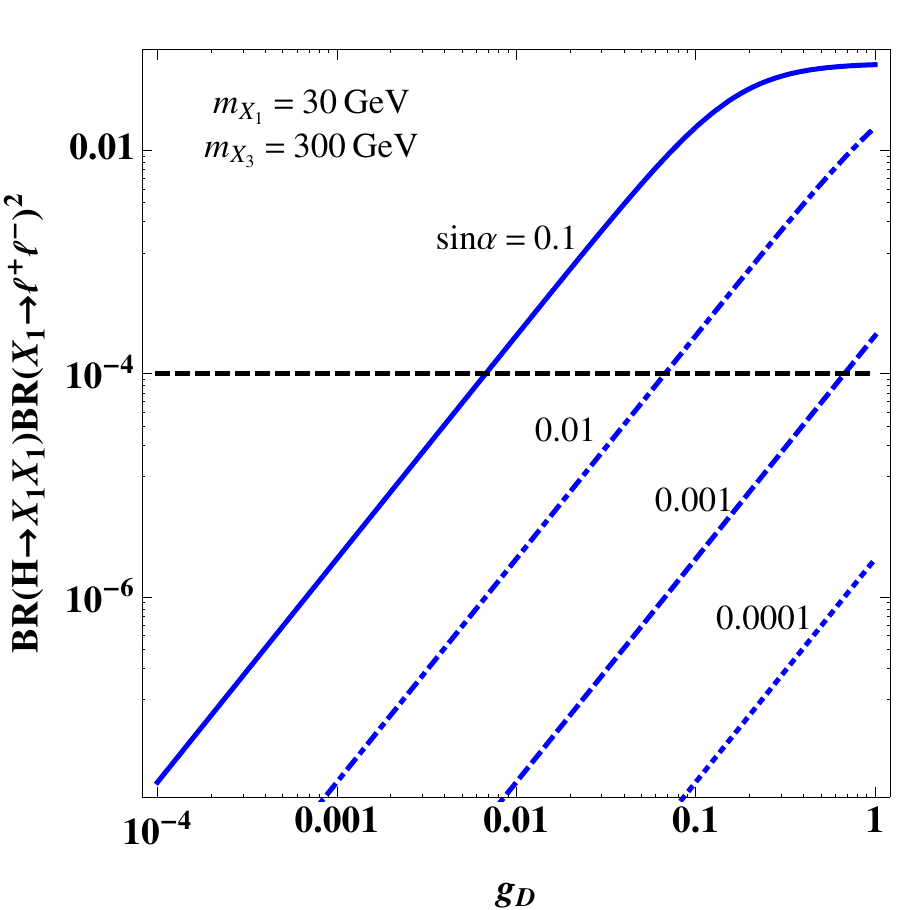} \
\includegraphics[width=80mm]{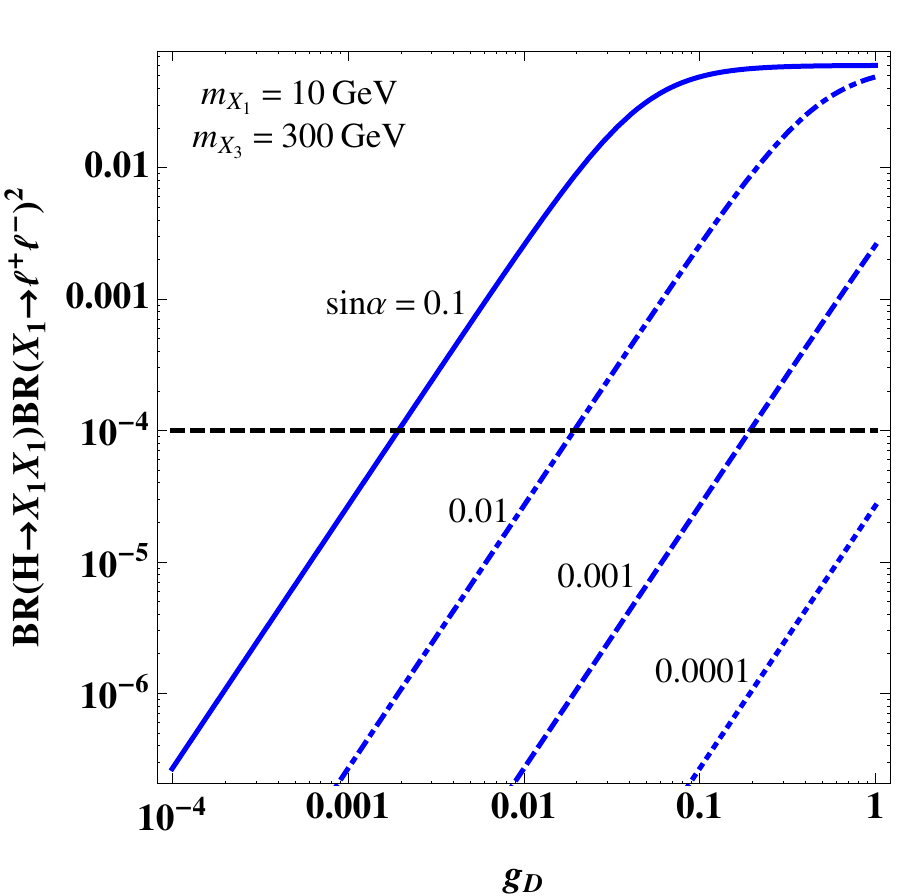}
\caption{Branching ratio for $h \to X_{1} X_{1} \to \ell^+ \ell^- \ell^+ \ell^-$ as a function of $SU(2)_D$ gauge coupling $g_D$. }   
\label{fig:Hdecay}\end{center}\end{figure}

Firstly we discuss constraints from the SM Higgs decay process, 
$h \to X_{1,2,3} X_{1,2,3} \to \ell^+ \ell^- \ell^+ \ell^-$ where $\ell$ denotes electron or muon.
This  multi-lepton decay channel is strongly constrained by the search for Higgs boson decaying into extra gauge boson which can decay into charged leptons~\cite{Aaboud:2018fvk} 
because of little  background.
The decay $h \to X_{1,2,3} X_{1,2,3}$ is induced via scalar mixing between the dark sector and 
the SM Higgs sector.  \\
For the scenario (1), we obtain the decay widths as
\begin{align}
\Gamma_{h \to X_1 X_1} & = \frac{{g_D}^4 \cos^2 \alpha}{8 \pi}  \frac{v_\phi^2}{m_{h}} \sqrt{1 - \frac{4 m_{X_1}^2}{m_{h}^2}} \left[ 2 + \frac{m_{h}^4}{4 m_{X_1}^4} \left( 1 - \frac{2 m_{X_1}^2 }{m_{h}^2} \right)^2 \right] , \\ 
\Gamma_{h \to X_2 X_2} & = \frac{{g_D}^4 \cos^2 \alpha}{8 \pi}  \frac{v_\phi^2}{m_{h}} \sqrt{1 - \frac{4 m_{X_2}^2}{m_{h}^2}} \left[ 2 + \frac{m_{h}^4}{4 m_{X_2}^4} \left( 1 - \frac{2 m_{X_2}^2 }{m_{h}^2} \right)^2 \right] . 
\end{align}
For the scenario (2), we also obtain the decay widths as
\begin{align}
\Gamma_{h \to X_3 X_3} & = \frac{{g_D}^4 \cos^2 \alpha'}{8 \pi}  \frac{v_{\phi'}^2}{m_{h}} \sqrt{1 - \frac{4 m_{X_3}^2}{m_{h}^2}} \left[ 2 + \frac{m_{h}^4}{4 m_{X_3}^4} \left( 1 - \frac{2 m_{X_3}^2 }{m_{h}^2} \right)^2 \right] , \\ 
\Gamma_{h \to X_2 X_2} & = \frac{{g_D}^4 \cos^2 \alpha'}{8 \pi}  \frac{v_{\phi'}^2}{m_{h}} \sqrt{1 - \frac{4 m_{X_2}^2}{m_{h}^2}} \left[ 2 + \frac{m_{h}^4}{4 m_{X_2}^4} \left( 1 - \frac{2 m_{X_2}^2 }{m_{h}^2} \right)^2 \right] . 
\end{align}
In Fig.~\ref{fig:Hdecay}, we show branching ratio (BR) for the process $h \to X_{1} X_{1} \to \ell^+ \ell^- \ell^+ \ell^-$ in the scenario (1) where we consider $2 m_{X_1} < m_h$ and $2 m_{X_{2,3}} < m_h$ for simplicity;
for the scenario (2) we obtain the same result replacing the role of $X_1$ and $X_3$.
We also show the upper limit on the BR as a dashed horizontal line.
It is then found that the scalar mixing angle and/or the gauge coupling $g_D$ should be suppressed 
when the SM Higgs can decay into dark gauge boson decaying into charged leptons.

\subsection{Scalar boson production}
\begin{figure}[tb]\begin{center}
\includegraphics[width=80mm]{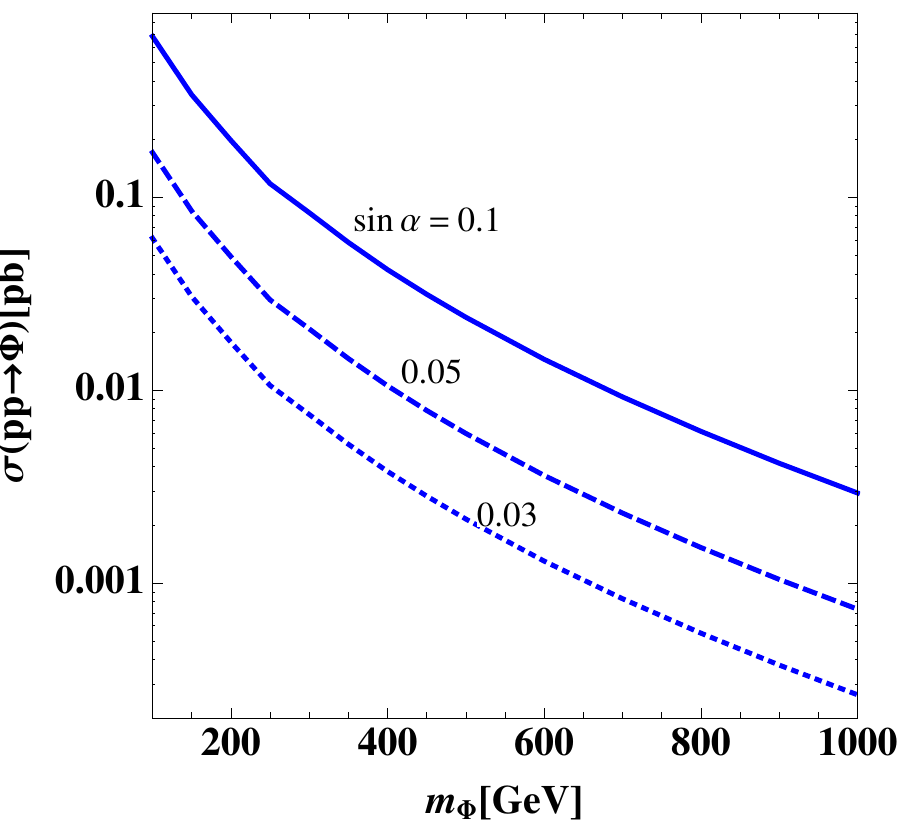} \
\caption{Cross section for $gg \to \Phi (\Phi = \Phi_1, \Phi'_1)$ process as a function of scalar mass with $\sqrt{s} = 14$ TeV.}   
\label{fig:1}\end{center}\end{figure}
Here we discuss $\Phi_1(\Phi'_1)$ production processes at the LHC. 
The scalar boson can be produced by gluon fusion process $gg \to \Phi_1(\Phi'_1)$ through 
the mixing with the SM Higgs boson parametrized by mixing angle $\alpha(\alpha')$.
The relevant effective interaction for the gluon fusion is written as~\cite{Gunion:1989we}
\begin{equation}
{\mathcal L}_{\phi gg} = \frac{\alpha_s}{16 \pi} \frac{\sin \alpha[\alpha']}{v} A_{1/2}(\tau_t) \phi G^a_{\mu \nu} G^{a \mu \nu}, 
 \end{equation}
 where $G^a_{\mu \nu}$ is the field strength for gluon and $A_{1/2}(\tau_t) = -\frac{1}{4} [\ln[(1+\sqrt{\tau_t})/(1-\sqrt{\tau_t})] - i \pi ]^2$ with $\tau_t = 4 m_{t}^2/m_\phi^2$.
We obtain this effective interaction from $\bar t t \Phi_1(\Phi'_1)$ coupling via the mixing effect where we take into account only top Yukawa coupling since the other contributions are subdominant.
In Fig.~\ref{fig:1}, we show the production cross section for scalar boson as a function of its mass with $\sqrt{s} = 14$ TeV adopting several values of $\sin \alpha(\alpha')$.
We find that a sizable scalar mixing is required to obtain observable cross section.
Thus we consider parameter region of $2 m_{X_1} > m_h$ in our discussion of collider physics 
since the scalar mixing is constrained for $2 m_{X_1} < m_h$ as shown in previous subsection.

\subsection{ Branching ratio of extra particles}
\begin{figure}[tb]\begin{center}
\includegraphics[width=80mm]{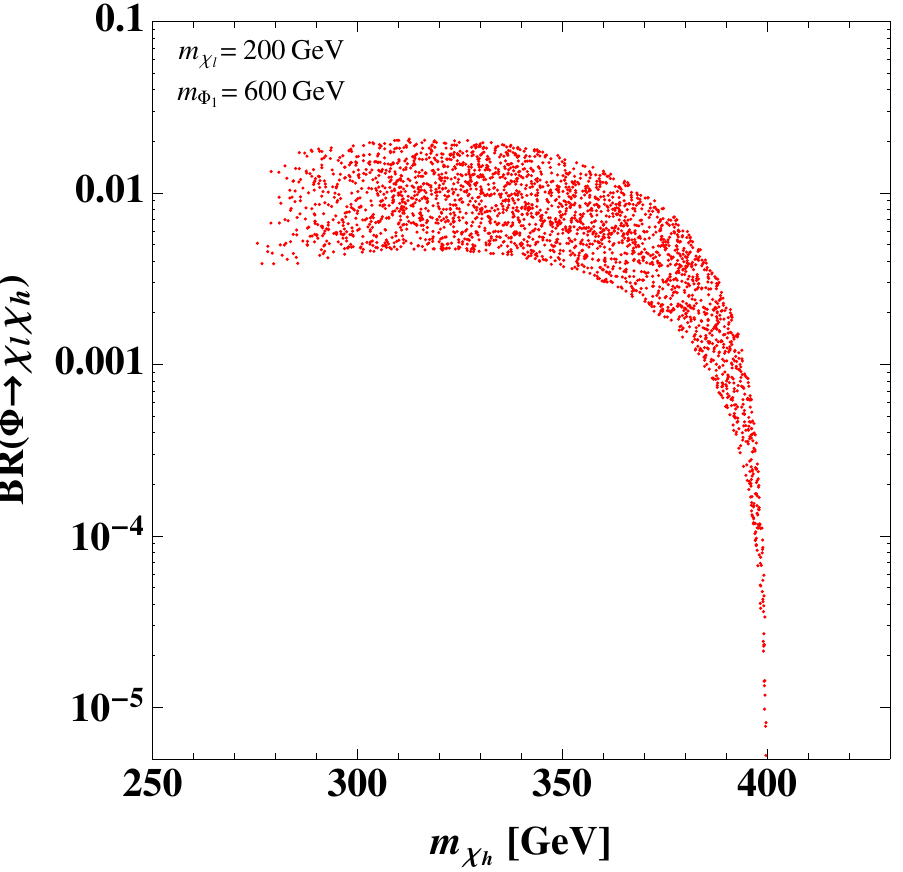} \
\includegraphics[width=80mm]{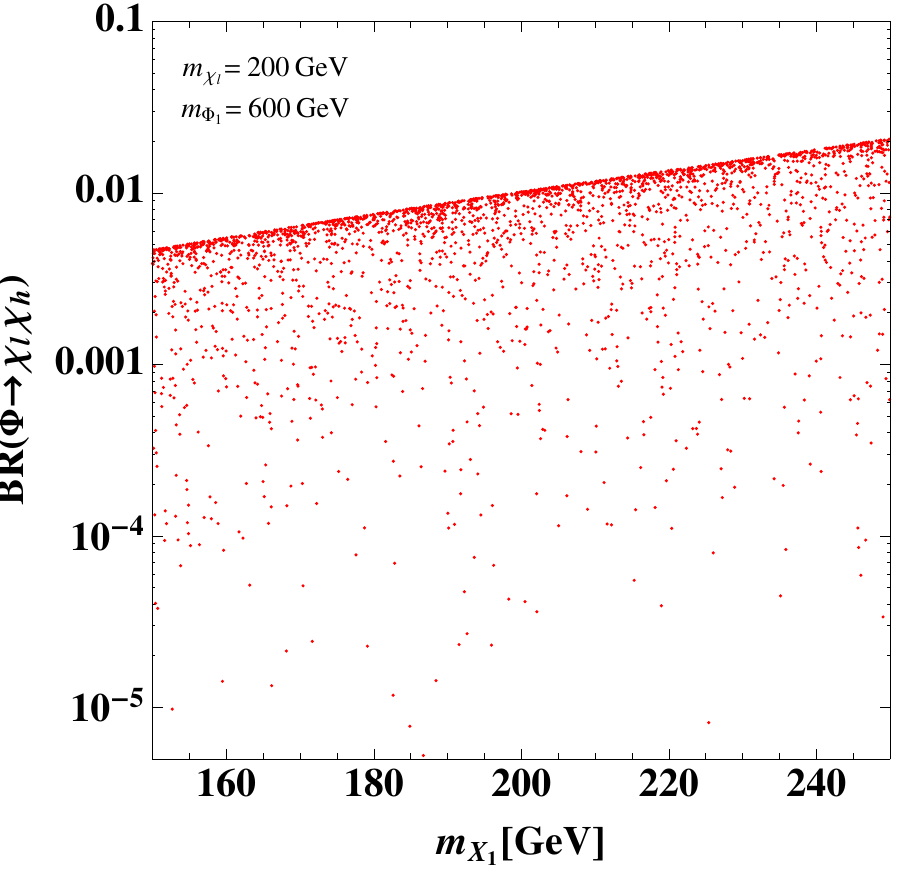}
\caption{Branching ratio for $\Phi_1 \to \chi_l \chi_h$ where final state includes both $\bar\chi_l \chi_h$ and $\bar \chi_h \chi_l$. }   
\label{fig:2}\end{center}\end{figure}
Here we estimate BRs of particles in dark sector.
The decay widths for the $\Phi_1[\Phi'_1] \to \chi_a \chi_b \ (a(b) = l, h)$ processes are given by
\begin{align}
& \Gamma_{\Phi_1[\Phi'_1] \to \chi_a \chi_b} = \frac{\left| Y_{\Phi_1[\Phi'_1]}^{\chi_a \chi_b} \right|^2}{8 \pi} m_{\Phi_1 [\Phi'_1]} \lambda^{\frac12}(m_{\Phi_1[\Phi'_1]}, m_{\chi_a}, m_{\chi_b} ) 
\left[ 1 - \frac{(m_{\chi_a} + m_{\chi_b})^2}{m_{\Phi_1[\Phi'_1]}^2} \right] \\
& Y_{\Phi_1}^{\chi_l \chi_l} = \frac{ y_{\chi \phi} \cos \alpha}{2} \cos 2 \theta_\chi, \ Y_{\Phi_1}^{\chi_h \chi_h} = -\frac{ y_{\chi \phi} \cos \alpha}{2} \cos 2 \theta_\chi, \ 
Y_{\Phi_1}^{\chi_l \chi_h} = - \frac{ y_{\chi \phi} \cos \alpha}{2} \sin 2 \theta_\chi, \nonumber \\
& Y_{\Phi'_1}^{\chi_l \chi_l} = \frac{ y_{\chi \phi'} \cos \alpha'}{2} \sin 2 \theta_\chi, \ Y_{\Phi'_1}^{\chi_h \chi_h} = -\frac{ y_{\chi \phi} \cos \alpha'}{2} \sin 2 \theta_\chi, \ 
Y_{\Phi_1}^{\chi_l \chi_h} = - \frac{ y_{\chi \phi} \cos \alpha'}{2} \cos 2 \theta_\chi, \nonumber \\
& \lambda(m_1,m_2,m_3) \equiv 1 + \frac{m_2^4}{m_1^4} + \frac{m_3^4}{m_1^4} -  \frac{2 m_2^2}{m_1^2} -  \frac{2 m_3^2}{m_1^2} -  \frac{2 m_2^2 m_3^2}{m_1^4}  \nonumber 
\end{align}
where $\Phi_1[\Phi'_1]$ is mass eigenstate which mixes with the SM Higgs as Eq.~\eqref{eq:massE3} and \eqref{eq:massE4}. 
The dark scalar bosons also decay into dark gauge bosons. 
For  the scenario (1), we obtain
\begin{align}
\Gamma_{\Phi_1 \to X_1 X_1} & = \frac{{g_D}^4 \cos^2 \alpha}{8 \pi}  \frac{v_\phi^2}{m_{\Phi_1}} \sqrt{1 - \frac{4 m_{X_1}^2}{m_{\Phi_1}^2}} \left[ 2 + \frac{m_{\Phi_1}^4}{4 m_{X_1}^4} \left( 1 - \frac{2 m_{X_1}^2 }{m_{\Phi_1}^2} \right)^2 \right] , \\ 
\Gamma_{\Phi_1 \to X_2 X_2} & = \frac{{g_D}^4 \cos^2 \alpha}{8 \pi}  \frac{v_\phi^2}{m_{\Phi_1}} \sqrt{1 - \frac{4 m_{X_2}^2}{m_{\Phi_1}^2}} \left[ 2 + \frac{m_{\Phi_1}^4}{4 m_{X_2}^4} \left( 1 - \frac{2 m_{X_2}^2 }{m_{\Phi_1}^2} \right)^2 \right] . 
\end{align}
For the scenario (2), we also obtain
\begin{align}
\Gamma_{\Phi'_1 \to X_3 X_3} & = \frac{{g_D}^4 \cos^2 \alpha'}{8 \pi}  \frac{v_{\phi'}^2}{m_{\Phi'_1}} \sqrt{1 - \frac{4 m_{X_3}^2}{m_{\Phi'_1}^2}} \left[ 2 + \frac{m_{\Phi'_1}^4}{4 m_{X_3}^4} \left( 1 - \frac{2 m_{X_3}^2 }{m_{\Phi'_1}^2} \right)^2 \right] , \\ 
\Gamma_{\Phi'_1 \to X_2 X_2} & = \frac{{g_D}^4 \cos^2 \alpha'}{8 \pi}  \frac{v_{\phi'}^2}{m_{\Phi'_1}} \sqrt{1 - \frac{4 m_{X_2}^2}{m_{\Phi'_1}^2}} \left[ 2 + \frac{m_{\Phi'_1}^4}{4 m_{X_2}^4} \left( 1 - \frac{2 m_{X_2}^2 }{m_{\Phi'_1}^2} \right)^2 \right] . 
\end{align}
In Fig.~\ref{fig:2}, we show BR for $\Phi_1 \to \chi_l \chi_h$ as functions of $m_{\chi_h}$ and 
$m_{X_1}$ in the scenario (1) where we have scanned coupling as $y_{\chi \phi}(g_D) \in [0.5, 2.5]
([1.5, 2.5])$ and fixed some parameters $\sin \alpha = 0.1$, $\theta_{\chi} = \pi/4$, $m_{\chi_l} = 200$ GeV, $m_{X_2} \simeq m_{X_3} = 500$ GeV and $m_{\Phi_1} = 600$ GeV.
We find the BR for $\Phi_1 \to \chi_l \chi_h$ is maximally $1.2 \times 10^{-2}$ and dominant decay 
mode is the $\Phi_1 \to X_1 X_1$ mode where the other modes are suppressed.
For the scenario (2), we obtain similar result by replacing $X_1$ and $X_3$ and the corresponding 
plot is omitted here.

\subsection{Signal at the LHC}

Here we discuss signature of our model at the LHC based on decay modes of extra scalar bosons
which are produced through gluon fusion process via scalar mixing. 
As we discuss in Sec.~\ref{sec:HiggsConst}, scalar mixing cannot be sizable when the SM Higgs decays into dark gauge bosons.
Thus dark gauge boson masses are assumed to be heavier than half of Higgs mass to realize observable signals from extra scalar production.
We summarize possible signature of the model in the following.

 {\bf (a)} {\it $\Phi_{1}[\Phi'_1] \to X_1 X_1 [X_3 X_3]$ decay mode}: 
For $m_{X_{1[3]}} < 2 m_{\chi_l}$, $X_{1[3]}$  dominantly decays into SM fermions induced by kinetic mixing. 
The BR of this decay chain of $\Phi_1(\Phi'_1)$ is dominant when it is kinematically allowed, $m_{\Phi_1[\Phi'_1]} > 2 m_{X_{1[3]}}$, and provide sizable cross section.
The most clear signal is four charged lepton final states which can be well tested at the LHC.
For $m_{X_{1[3]}} > 2 m_{\chi_l}$, $X_{1[3]}$ dominantly decay into DM since SM fermion mode is suppressed by small kinetic mixing.
In this case, the final state becomes transverse missing energy and we need additional jet/photon for tagging.

{\bf (b)} {\it $\Phi_{1}[\Phi'_1] \to X_2 X_2$ decay mode}: 
For $m_{X_{2}} <  m_{\chi_l} + m_{\chi_h}$,  our signal is eight SM fermions coming from decay chain of $X_2 \to X_1 X_3 (X_{1,3} \to \bar f_{SM} f_{SM})$. 
The BR of this decay mode of $\Phi_1(\Phi'_1)$ can be sizable when it is kinematically allowed and masses among dark gauge bosons are not hierarchical.
For $m_{X_{2}} >  m_{\chi_l} + m_{\chi_h}$, $X_2$ dominantly decays into $\bar \chi_h \chi_l(\bar \chi_l \chi_h)$.
Then $\chi_h$ decays as $\chi_h \to X_{1[3]}^{(*)} \chi_l$ where dark gauge boson is off-shell or on-shell depending on mass hierarchy.
In this case, we obtain signal of four SM fermions with missing transverse momentum from the decay chain. 

{\bf (c)} {\it  $\Phi_{1}[\Phi'_1] \to \chi_l \chi_h$ decay mode}: 
In this case our signal is SM fermion pair with missing transverse momentum coming from decay chain of $\chi_h \to X_{1,3} \chi_l (X_{1,3} \to \bar f_{SM} f_{SM})$. 
For $m_{\Phi_1[\Phi'_1]} > 2 m_{X_{1[3]}}$, the BR for the decay mode is small as we discussed above but
we can still obtain $\sim$ 30 events when $\sigma(gg \to \Phi_{1}[\Phi'_1]) = 10$ fb, $BR(\Phi_1[\Phi'_1] \to \chi_l \chi_h) \sim  10^{-2}$ and integrated luminosity is $L = 300$ fb$^{-1}$.
On the other hand, for $m_{\Phi_1[\Phi'_1]} < 2 m_{X_{1[3]}}$, the BR of this decay chain is dominant and we can obtain cross section as large as $\Phi_1[\Phi'_1]$ production cross section.

We indicate dominant decay mode of $\Phi_1$ for some mass relations in scenario (1)
where we can obtain similar results in scenario (3) interchanging the role of $\{\Phi_1, X_1 \}$ and $\{\Phi'_1 X_3 \}$.
In Table~\ref{tab:3}, we show cross sections of signal processes for some benchmark points (BPs) where these parameters can be consistent with relic density of DM.
We thus find that signal cross sections can be sizable when scalar mixing is not too small, 
and signatures of our model at the collider experiments depend on mass relation in dark sector.
Combining analysis of these processes we can test our model at the LHC and the HL-LHC.
Further analysis with detailed simulation is beyond the scope of this work and it is left for future work.

 \begin{widetext}
\begin{center} 
\begin{table}[t]
\begin{tabular}{|c||c|c|c|}\hline\hline  
~Mass relation~ &  $m_{\Phi_1} > 2 m_{X_1}$, $m_{X_3} > m_{X_1}$ & $m_{\Phi_1} > 2 m_{X_1}$, $m_{X_1} \sim m_{X_3}$ & $m_{\Phi_1} < 2 m_{X_{1,3}}$, $m_{\Phi_1} > m_{\chi_l}+ m_{\chi_h}$  \\\hline 
& $X_1 X_1$ & $X_1 X_1$, $X_2 X_2$, $X_3 X_3$ & $\bar \chi_l \chi_h (\bar \chi_h \chi_l)$ \\ \hline
\end{tabular}
\caption{Dominant decay mode of $\Phi_1$ for some mass relations.}
\label{tab:2}
\end{table}
\end{center}
\end{widetext}

 \begin{widetext}
\begin{center} 
\begin{table}[t]
\begin{tabular}{|c||c|cc|}\hline\hline  
BP &  parameters & ~~final states~~ & ~~$\sigma BR$~~ \\\hline 
1 &  $\{m_\Phi, m_{\chi_l}, m_{\chi_h}, m_{X_1}, m_{X_3} \} = \{800, 300, 401, 200, 500\} $ [GeV] & $2 \bar f_{SM} f_{SM}$ & $\sim 6$ [fb]  \\
   &  $\{g_D, y_{\chi \phi}, y_{\chi \phi'}, \sin \alpha \} = \{0.4, 0.1, 0.1, 0.1 \}$ &  $\bar  \chi_l  \chi_l   \bar f_{SM} f_{SM}$  &  $\sim 0.01$ [fb] \\  \hline
2 &  $\{m_\Phi, m_{\chi_l}, m_{\chi_h}, m_{X_1}, m_{X_2} \} = \{600, 200, 335, 200, 200\} $ [GeV] &  $ 2  \bar f_{SM} f_{SM}$  &  $\sim 12.8$ [fb] \\
   &  $\{g_D, y_{\chi \phi}, y_{\chi \phi'}, \sin \alpha \} = \{0.7, 0.1, 0.1, 0.1 \}$ & $4 \bar f_{SM} f_{SM}$ & $\sim 1.6$ [fb] \\  \hline
3 &  $\{m_\Phi, m_{\chi_l}, m_{\chi_h}, m_{X_1}, m_{X_2} \} = \{300, 75, 125, 150, 200\} $ [GeV] &  $\bar  \chi_l  \chi_l   \bar f_{SM} f_{SM}$  &  $\sim 83$ [fb] \\
   &  $\{g_D, y_{\chi \phi}, y_{\chi \phi'}, \sin \alpha \} = \{0.7, 0.2, 0.2, 0.1 \}$ & & \\  \hline   
\end{tabular}
\caption{Cross sections for signal processes in some benchmark points (BPs) for scenario (1) at the LHC 14 TeV. The numbers in front of $\bar f_{SM} f_{SM}$ indicate number of SM fermion anti-fermion pair in final states.}
\label{tab:3}
\end{table}
\end{center}
\end{widetext}

\section{ Summary and discussions}

We have discussed a model of dark sector described by $SU(2)_D$ gauge symmetry 
in which two triplet real scalar fields and one doublet Dirac fermion are introduced.
In our scenario, $SU(2)_D$ symmetry is broken to discrete $Z_2$ symmetry by VEVs of 
two triplet scalar fields.
Then remaining $Z_2$ symmetry guarantees stability of DM candidate which is the lighter 
component from doublet fermion $\chi_l$.
In the gauge sector, we consider kinetic mixing term between $SU(2)_D$ and $U(1)_Y$ which 
is assumed to be generated via 5-dimensional operators.
{Then we have investigated dark gauge sector  
which provides three massive dark gauge bosons $X_{1,2,3}$, two of which can mix with SM gauge boson 
via the kinetic mixings. }

We have estimated relic density of our DM candidate where the observed value is explained 
via gauge interactions in dark sector with kinetic mixing effect as a portal to the SM sector.
Then we have explored parameter region satisfying observed relic density. 
We have found that the relic density is explained by the process, 
$\chi_l \bar \chi_l \to X_{1,2,3}X_{1,2,3}$, in large parameter region 
while we need fine tuning to obtain resonant enhancement for the process,  
$\chi_l \bar \chi_l \to f_{SM} \bar f_{SM}$, via dark gauge boson exchange with kinetic mixing. 

Implications to collider physics have been discussed such as $h \to Z'Z'$ decay,  
and extra scalar production and its possible signals at the LHC.
We have found that the constraint from $h \to Z' Z'$ branching ratio restricts scalar mixing 
with the SM Higgs and $SU(2)_D$ gauge coupling severely when the mode is kinematically allowed.
Extra scalar boson can be produced by gluon fusion process through scalar mixing associated with the SM Higgs. 
For extra scalar production, we obtain some specific signatures depending on mass relation of dark sector particles.

\section*{Acknowledgments}
\vspace{0.5cm}
The work is supported in part by KIAS Individual Grants, Grant No. PG021403 (PK) and {No. PG054702 (TN) }at 
Korea Institute for Advanced Study, and by National Research Foundation of Korea (NRF) Grant 
No. NRF-2019R1A2C3005009 (PK), funded by the Korea government (MSIT).
This research was supported by an appointment to the JRG Program at the APCTP through the Science and Technology Promotion Fund and Lottery Fund of the Korean Government. This was also supported by the Korean Local Governments - Gyeongsangbuk-do Province and Pohang City (H.O.). 
H.O.~is sincerely grateful for KIAS and the members who provide me huge hospitality during my stay.



\end{document}